\documentclass[conference]{IEEEtran}
\IEEEoverridecommandlockouts
\usepackage{cite}
\usepackage{amsmath,amssymb,amsfonts}
\usepackage{algorithmic}
\usepackage{graphicx}
\usepackage{textcomp}
\usepackage{xcolor}
\usepackage{tikz}
\usepackage{balance}
\usetikzlibrary{positioning, arrows.meta, fit, backgrounds}

\usetikzlibrary{
positioning,
arrows.meta,
fit,
backgrounds,
calc,
shapes.geometric,
shadows
}

\definecolor{deepblue}{HTML}{1E3A8A}
\definecolor{softblue}{HTML}{DBEAFE}
\definecolor{softcyan}{HTML}{CFFAFE}
\definecolor{softgreen}{HTML}{DCFCE7}
\definecolor{softorange}{HTML}{FFEDD5}
\definecolor{softpurple}{HTML}{EDE9FE}
\definecolor{softred}{HTML}{FEE2E2}
\definecolor{softgray}{HTML}{F8FAFC}
\definecolor{linegray}{HTML}{334155}

\def\BibTeX{{\rm B\kern-.05em{\sc i\kern-.025em b}\kern-.08em
    T\kern-.1667em\lower.7ex\hbox{E}\kern-.125emX}}

\begin{document}

\title{Trustworthy Self-Composable Big-Data-as-a-Service: An LLM-Orchestrated Multi-Agent Framework for Automated Data Engineering, AutoML, MLOps Deployment, and Drift-Aware Lifecycle Optimization\\

}

\author{
\IEEEauthorblockN{
Aueaphum Aueawatthanaphisut\IEEEauthorrefmark{1},
Badri Raj Lamichhane\IEEEauthorrefmark{2}
}
\IEEEauthorblockA{
School of Information, Computer, and Communication Technology\\
Sirindhorn International Institute of Technology, Thammasat University\\
Pathum Thani, Thailand\\
\IEEEauthorrefmark{1}aueawatth.aue@gmail.com,
\IEEEauthorrefmark{2}d6622300231@g.siit.tu.ac.th
}
}

\maketitle

\begin{abstract}
Big-Data-as-a-Service (BDaaS) platforms require reliable automation across data ingestion, cleaning, feature engineering, model development, deployment, and post-deployment monitoring. However, existing LLM-based data science agents and AutoML systems mainly focus on isolated workflow stages, leaving limited support for lifecycle-level orchestration, artifact governance, human oversight, and drift-aware adaptation. This paper proposes a trustworthy self-composable BDaaS framework based on LLM-orchestrated multi-agent collaboration. The proposed architecture decomposes the BDaaS lifecycle into specialized agents for data ingestion, data cleaning, feature engineering, AutoML training, model evaluation, MLOps deployment, monitoring, and drift detection. A central LLM orchestration layer coordinates agent execution, validates intermediate outputs, manages workflow context, and enables dynamic workflow composition. The framework also incorporates shared artifact governance, reproducibility support, human-in-the-loop checkpoints, and drift-aware feedback loops. A prototype-based evaluation is conducted using controlled tabular benchmark datasets with missing values, categorical variables, outliers, class imbalance, and simulated covariate drift. Compared with manual ML, AutoML-only, and single-agent LLM baselines, the proposed multi-agent BDaaS pipeline achieves competitive predictive performance while improving lifecycle-level reliability, including workflow completion, artifact traceability, deployment readiness, reproducibility, and drift recovery. The results suggest that LLM-orchestrated multi-agent systems can extend conventional AutoML toward trustworthy, adaptive, and production-oriented BDaaS lifecycle automation.
\end{abstract}


\section{Introduction}

The increasing volume and complexity of data-intensive applications have created a growing demand for scalable, automated, and reliable Big-Data-as-a-Service (BDaaS) platforms. In practical machine learning workflows, organizations must repeatedly perform data ingestion, cleaning, feature engineering, model development, deployment, and post-deployment monitoring. Although cloud and big-data infrastructures reduce computational complexity, the end-to-end lifecycle still requires substantial human expertise to design pipelines, validate data quality, tune models, prepare deployment artifacts, and respond to drift or performance degradation.

Recent advances in large language models (LLMs) have introduced new opportunities for automating data-centric and machine learning workflows. LLM-based agents have demonstrated strong capabilities in reasoning, planning, code generation, data analysis, visualization, and machine learning experimentation. Prior studies such as AutoML-Agent, Data Interpreter, DS-Agent, MLAgentBench, and MLE-bench show that LLM agents can automate or assist multiple stages of data science and machine learning development \cite{b1,b2,b10,b11,b12}. Other works further demonstrate the use of LLMs for data cleaning, feature engineering, and visualization \cite{b3,b4,b5,b6,b15,b16}. These studies indicate that LLM agents can serve not only as coding assistants but also as coordinators for multi-step analytical workflows.

However, most existing approaches focus on isolated stages such as AutoML, data cleaning, feature engineering, visualization, or experimentation. In real BDaaS environments, these stages are tightly connected: changes in data schema can affect feature engineering, model updates may require deployment reconfiguration, and post-deployment drift may trigger retraining or rollback. Moreover, trustworthy automation remains challenging because production systems require reproducibility, intermediate validation, human oversight, artifact governance, and lifecycle-level monitoring \cite{b18,b19,b20}. Without such coordination, automated workflows may improve local efficiency but introduce risks in reliability, maintainability, and governance.

To address these limitations, this paper proposes a trustworthy self-composable BDaaS framework based on LLM-orchestrated multi-agent collaboration. The framework decomposes the BDaaS lifecycle into specialized agents for data ingestion, data cleaning, feature engineering, AutoML, model evaluation, deployment preparation, monitoring, and drift-aware optimization. An LLM orchestration layer coordinates these agents, validates intermediate outputs, manages workflow context, and supports dynamic workflow composition according to user requirements, dataset characteristics, and operational constraints.

The main contributions of this paper are threefold. First, we present a unified LLM-orchestrated multi-agent architecture for self-composable BDaaS. Second, we define the functional roles of agents across automated data engineering, AutoML, MLOps deployment, and lifecycle monitoring. Third, we introduce trustworthiness mechanisms, including validation checkpoints, human-in-the-loop control, artifact governance, reproducibility support, and drift-aware feedback loops. The proposed framework aims to reduce manual workload while improving adaptability, transparency, and reliability in end-to-end big-data and machine learning service environments.

\section{Related Work}

Recent advances in large language models (LLMs) have enabled agentic systems that can automate data science, machine learning experimentation, and workflow generation. AutoML-Agent demonstrates a multi-agent LLM framework for full-pipeline AutoML, covering preprocessing, model selection, training, and evaluation \cite{b1}. Data Interpreter shows that LLM agents can solve data science tasks through reasoning, code generation, and tool use \cite{b2}. Similarly, DS-Agent introduces case-based reasoning to improve automated data science workflows by reusing prior task experiences \cite{b10}. These works demonstrate the potential of LLM-based agents for automating analytical workflows, but they mainly focus on experimentation and model development rather than complete Big-Data-as-a-Service (BDaaS) lifecycle management.

Several studies have explored LLMs for specific data engineering tasks. LIDA uses LLMs to generate data visualizations and infographics from datasets \cite{b3}, while AutoDCWorkflow investigates LLM-based data cleaning workflow generation and benchmarking \cite{b4}. Bendinelli \textit{et al.} further examine LLM agents for cleaning tabular machine learning datasets \cite{b5}. In addition, systematic studies on data cleaning confirm that data quality issues such as missing values, inconsistent formats, and outliers strongly affect machine learning reliability \cite{b8}. Automated feature engineering has also been studied through LLM fine-tuning and context-aware feature construction, as shown by Hirose \textit{et al.} \cite{b6} and CAAFE \cite{b15}. Broader surveys on automated data processing and data-centric AI further emphasize the importance of scalable preprocessing, feature engineering, and data lifecycle management \cite{b7,b9}. However, these methods are generally designed as individual components and do not fully integrate data ingestion, cleaning, feature engineering, model development, deployment, and monitoring into a unified BDaaS framework.

Benchmarking and improving LLM-based machine learning agents is another active research direction. MLAgentBench evaluates language agents on machine learning experimentation tasks \cite{b11}, while MLE-bench provides more realistic machine learning engineering challenges involving implementation, debugging, and optimization \cite{b12}. BudgetMLAgent studies cost-effective multi-agent automation for machine learning tasks \cite{b13}, and SELA improves AutoML agents through tree-search-based reasoning \cite{b14}. AIDE also explores AI-driven code exploration for generating and refining executable* workflows \cite{b17}. These studies show strong progress in autonomous ML experimentation, but most evaluations remain development-oriented and do not sufficiently address production concerns such as artifact versioning, deployment readiness, rollback, monitoring, and lifecycle governance.

MLOps research highlights the importance of reliable deployment and continuous maintenance of machine learning systems. Kreuzberger \textit{et al.} define MLOps as the integration of machine learning, DevOps, and data engineering to support continuous integration, delivery, monitoring, and governance \cite{b18}. Mallick \textit{et al.} study data drift mitigation in large-scale online services, showing that deployed models must adapt to changing data distributions \cite{b19}. Sculley \textit{et al.} further warn that unmanaged machine learning pipelines can create hidden technical debt and reduce maintainability \cite{b20}. These works motivate the need for trustworthy lifecycle management in BDaaS platforms. Nevertheless, existing approaches remain fragmented across LLM agents, data engineering, AutoML, and MLOps. This paper addresses this gap by proposing an LLM-orchestrated multi-agent BDaaS framework that unifies automated data engineering, AutoML, deployment preparation, artifact governance, human oversight, and drift-aware lifecycle optimization within a single self-composable architecture.

\section{Methodology}

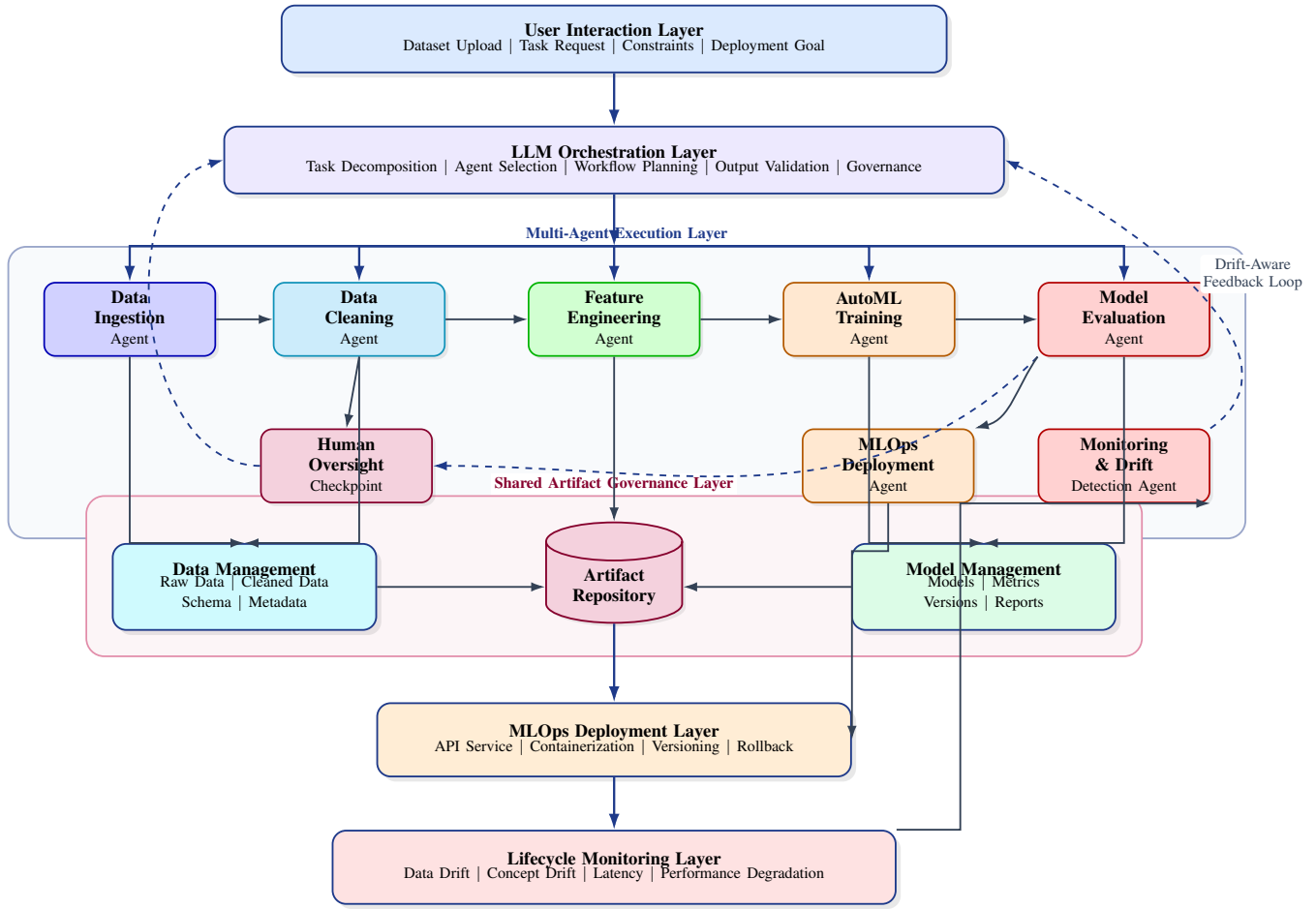
\begin{figure*}[t]
\centering
\resizebox{\textwidth}{!}{
\begin{tikzpicture}[
font=\footnotesize,
x=1cm,y=1cm,
layer/.style={
rectangle,
rounded corners=6pt,
draw=deepblue,
line width=0.8pt,
fill=#1,
align=center,
minimum height=1.05cm,
text width=8.8cm,
drop shadow={shadow xshift=0.8mm, shadow yshift=-0.8mm, opacity=0.18}
},
agent/.style={
rectangle,
rounded corners=5pt,
draw=#1!70!black,
line width=0.75pt,
fill=#1!18,
align=center,
minimum height=1.15cm,
text width=2.45cm,
drop shadow={shadow xshift=0.55mm, shadow yshift=-0.55mm, opacity=0.16}
},
store/.style={
cylinder,
shape border rotate=90,
aspect=0.28,
draw=#1!70!black,
line width=0.8pt,
fill=#1!18,
align=center,
minimum height=1.35cm,
minimum width=1.8cm,
text width=1.9cm,
drop shadow={shadow xshift=0.55mm, shadow yshift=-0.55mm, opacity=0.16}
},
arrow/.style={
-{Latex[length=2.2mm,width=1.5mm]},
line width=0.8pt,
draw=linegray
},
mainarrow/.style={
-{Latex[length=2.5mm,width=1.7mm]},
line width=1.0pt,
draw=deepblue
},
dashedarrow/.style={
-{Latex[length=2.2mm,width=1.5mm]},
line width=0.8pt,
draw=deepblue,
dashed
},
smalllabel/.style={
font=\scriptsize,
align=center,
text=linegray,
fill=white,
inner sep=1.5pt,
rounded corners=2pt
}
]

% ======================================================
% Top Layers
% ======================================================
\node[layer=softblue, text width=10.2cm] (user) at (0,0) {
\textbf{User Interaction Layer}\\[-1mm]
\scriptsize Dataset Upload $\vert$ Task Request $\vert$ Constraints $\vert$ Deployment Goal
};

\node[layer=softpurple, text width=12.0cm] (llm) at (0,-1.9) {
\textbf{LLM Orchestration Layer}\\[-1mm]
\scriptsize Task Decomposition $\vert$ Agent Selection $\vert$ Workflow Planning $\vert$ Output Validation $\vert$ Governance
};

% ======================================================
% Main Agent Row
% ======================================================
\node[agent=blue] (ingest) at (-7.6,-4.4){\textbf{Data}\\\textbf{Ingestion}\\\scriptsize Agent};
\node[agent=cyan]   (clean)   at (-4.0,-4.4) {\textbf{Data}\\\textbf{Cleaning}\\\scriptsize Agent};
\node[agent=green]  (feature) at ( 0.0,-4.4) {\textbf{Feature}\\\textbf{Engineering}\\\scriptsize Agent};
\node[agent=orange] (automl)  at ( 4.0,-4.4) {\textbf{AutoML}\\\textbf{Training}\\\scriptsize Agent};
\node[agent=red]    (eval)    at ( 8.0,-4.4) {\textbf{Model}\\\textbf{Evaluation}\\\scriptsize Agent};

% ======================================================
% Secondary Agent Row
% ======================================================
\node[agent=purple] (human)       at (-4.2,-6.7) {\textbf{Human}\\\textbf{Oversight}\\\scriptsize Checkpoint};
\node[agent=orange] (deployagent) at ( 4.3,-6.7) {\textbf{MLOps}\\\textbf{Deployment}\\\scriptsize Agent};
\node[agent=red]    (monitoragent)at ( 8.0,-6.7) {\textbf{Monitoring}\\\textbf{\& Drift}\\\scriptsize Detection Agent};

% ======================================================
% Repository and Management Layer
% ======================================================
\node[layer=softcyan, text width=3.9cm, minimum height=1.35cm] (datarepo) at (-5.8,-8.6) {
\textbf{Data Management}\\[-1mm]
\scriptsize Raw Data $\vert$ Cleaned Data\\
\scriptsize Schema $\vert$ Metadata
};

\node[store=purple] (repo) at (0,-8.6) {
\textbf{Artifact}\\
\textbf{Repository}
};

\node[layer=softgreen, text width=3.9cm, minimum height=1.35cm] (modelrepo) at (5.8,-8.6) {
\textbf{Model Management}\\[-1mm]
\scriptsize Models $\vert$ Metrics\\
\scriptsize Versions $\vert$ Reports
};

% ======================================================
% Deployment / Monitoring Layers
% ======================================================
\node[layer=softorange, text width=7.2cm, minimum height=1.15cm] (deploy) at (0,-11.0) {
\textbf{MLOps Deployment Layer}\\[-1mm]
\scriptsize API Service $\vert$ Containerization $\vert$ Versioning $\vert$ Rollback
};

\node[layer=softred, text width=8.6cm, minimum height=1.15cm] (monitor) at (0,-13.0) {
\textbf{Lifecycle Monitoring Layer}\\[-1mm]
\scriptsize Data Drift $\vert$ Concept Drift $\vert$ Latency $\vert$ Performance Degradation
};

% ======================================================
% Top Flow
% ======================================================
\draw[mainarrow] (user) -- (llm);

% fan-out from llm
\coordinate (branch) at (0,-3.25);
\draw[mainarrow] (llm.south) -- (branch);
\draw[mainarrow] (branch) -| (ingest.north);
\draw[mainarrow] (branch) -| (clean.north);
\draw[mainarrow] (branch) -- (feature.north);
\draw[mainarrow] (branch) -| (automl.north);
\draw[mainarrow] (branch) -| (eval.north);

% main pipeline
\draw[arrow] (ingest) -- (clean);
\draw[arrow] (clean) -- (feature);
\draw[arrow] (feature) -- (automl);
\draw[arrow] (automl) -- (eval);

% ======================================================
% Human Oversight / Validation
% ======================================================
\draw[arrow] (clean.south) -- (human.north);
\draw[dashedarrow] (eval.south west) to[out=-135,in=0] (human.east);

\draw[dashedarrow] (human.west) to[out=180,in=-165]
(llm.west);

% ======================================================
% Artifact / Data / Model Logging
% ======================================================
\draw[arrow] (ingest.south) |- (datarepo.north);
\draw[arrow] (clean.south) |- (datarepo.north);
\draw[arrow] (feature.south) -- (repo.north);
\draw[arrow] (automl.south) |- (modelrepo.north);
\draw[arrow] (eval.south) |- (modelrepo.north);

\draw[arrow] (datarepo.east) -- (repo.west);
\draw[arrow] (modelrepo.west) -- (repo.east);

% ======================================================
% Deployment / Monitoring Path
% ======================================================
% \draw[arrow] (eval.south) -- (monitoragent.north);

\draw[arrow] (eval.south west) to[out=-125,in=15] (deployagent.north east);

\draw[arrow] (deployagent.south) -- ++(0,-0.75) -| (deploy.east);
\draw[mainarrow] (repo) -- (deploy);
\draw[mainarrow] (deploy) -- (monitor);

\draw[arrow] (monitor.north east) -- ++(1.0,0) |- (monitoragent.south east);

\draw[dashedarrow] (monitoragent.north east) to[out=35,in=-25]
node[smalllabel, right, xshift=1mm, yshift=2mm] {Drift-Aware\\Feedback Loop}
(llm.east);
% ======================================================
% Background Boxes
% ======================================================
\begin{pgfonlayer}{background}

\node[
draw=deepblue!45,
line width=0.8pt,
rounded corners=8pt,
fill=softgray,
fit=(ingest)(clean)(feature)(automl)(eval)(human)(deployagent)(monitoragent),
inner sep=0.55cm
] (agentbox) {};

\node[
font=\scriptsize\bfseries,
text=deepblue,
fill=white,
inner sep=1.5pt
] at ($(agentbox.north)+(0,0.18)$) {Multi-Agent Execution Layer};

\node[
draw=purple!45,
line width=0.75pt,
rounded corners=8pt,
fill=purple!4,
fit=(datarepo)(repo)(modelrepo),
inner sep=0.40cm
] (govbox) {};

\node[
font=\scriptsize\bfseries,
text=purple!70!black,
fill=white,
inner sep=1.5pt
] at ($(govbox.north)+(0,0.18)$) {Shared Artifact Governance Layer};

\end{pgfonlayer}

\end{tikzpicture}
}
\caption{Proposed Q1-style LLM-orchestrated multi-agent architecture for trustworthy self-composable Big-Data-as-a-Service, integrating automated data engineering, AutoML, MLOps deployment, artifact governance, human oversight, and drift-aware lifecycle optimization.}
\label{fig:architecture}
\end{figure*}

This section describes the proposed trustworthy self-composable Big-Data-as-a-Service (BDaaS) framework. As illustrated in Fig.~\ref{fig:architecture}, the framework is designed as an LLM-orchestrated multi-agent architecture that coordinates the full lifecycle of data engineering, AutoML, deployment, and post-deployment monitoring. The core principle is to decompose a complex BDaaS workflow into specialized agents while maintaining centralized orchestration, shared artifact governance, human oversight, and drift-aware feedback control.

The framework in Fig.1 begins at the user interaction layer, where users submit high-level requests such as dataset upload, task definition, constraints, and deployment goals. These inputs are converted into structured task specifications and passed to the LLM orchestration layer. The orchestrator analyzes the request, decomposes it into executable* subtasks, selects suitable* agents, defines execution order, and validates intermediate outputs. Instead of treating data science automation as a fixed pipeline, the orchestrator dynamically composes workflows according to dataset characteristics, task requirements, model performance, and operational constraints.

The multi-agent execution layer contains specialized agents responsible for the major stages of the BDaaS lifecycle. The Data Ingestion Agent profiles heterogeneous data sources and extracts schema-level metadata. The Data Cleaning Agent identifies missing values, duplicated records, inconsistent formats, outliers, and invalid entries. The Feature Engineering Agent performs transformation, encoding, scaling, and feature construction using both statistical signals and semantic context. The AutoML Agent conducts model selection, hyperparameter tuning, training, and validation. The Model Evaluation Agent independently verifies whether candidate models satisfy predefined performance and reliability criteria. The MLOps Deployment Agent packages approved models into deployable services, including preprocessing pipelines, model files, API interfaces, container configurations, and versioned deployment artifacts. Finally, the Monitoring and Drift Detection Agent observes deployed services and reports data drift, concept drift, latency changes, and performance degradation to the orchestrator.

A key component of the proposed framework is the shared artifact governance layer. This layer stores raw and cleaned data, metadata, feature configurations, trained models, evaluation reports, deployment artifacts, and version histories. By maintaining these artifacts in a unified repository, the framework supports reproducibility, traceability, rollback, and lifecycle-level governance. This design reduces the risk of hidden technical debt by ensuring that each stage of the automated workflow can be inspected, reproduced, and revised when necessary.

Trustworthiness is embedded through validation checkpoints and human-in-the-loop control. Each agent returns structured outputs to the orchestrator, which checks whether the result satisfies the requirements of the next stage. Critical operations, such as data cleaning decisions, feature removal, model approval, deployment release, and drift response, can be routed to the Human Oversight Agent before execution. This mechanism allows the system to balance automation efficiency with human controllability, making the framework more suitable* for high-impact production environments.

After deployment, the lifecycle monitoring layer continuously evaluates service behavior using operational and statistical indicators, including request latency, prediction distribution, input feature distribution, error rate, and model performance when ground-truth labels are available. When drift or degradation is detected, the Monitoring and Drift Detection Agent sends feedback to the LLM orchestrator. The orchestrator can then trigger corrective actions such as alert generation, retraining, model comparison, preprocessing revision, redeployment, or rollback to a stable* version. In this way, the workflow does not terminate at model deployment but remains adaptive throughout the service lifecycle.

Overall, the proposed methodology provides a unified architecture for self-composable BDaaS by combining LLM-based orchestration, modular agent execution, artifact governance, MLOps deployment, human oversight, and drift-aware feedback loops. This design aims to reduce manual workload while improving transparency, adaptability, reproducibility, and operational reliability in end-to-end big-data and machine learning service environments.

\section{Experimental Design}

\begin{figure*}[t]
\centering
\includegraphics[width=\textwidth]{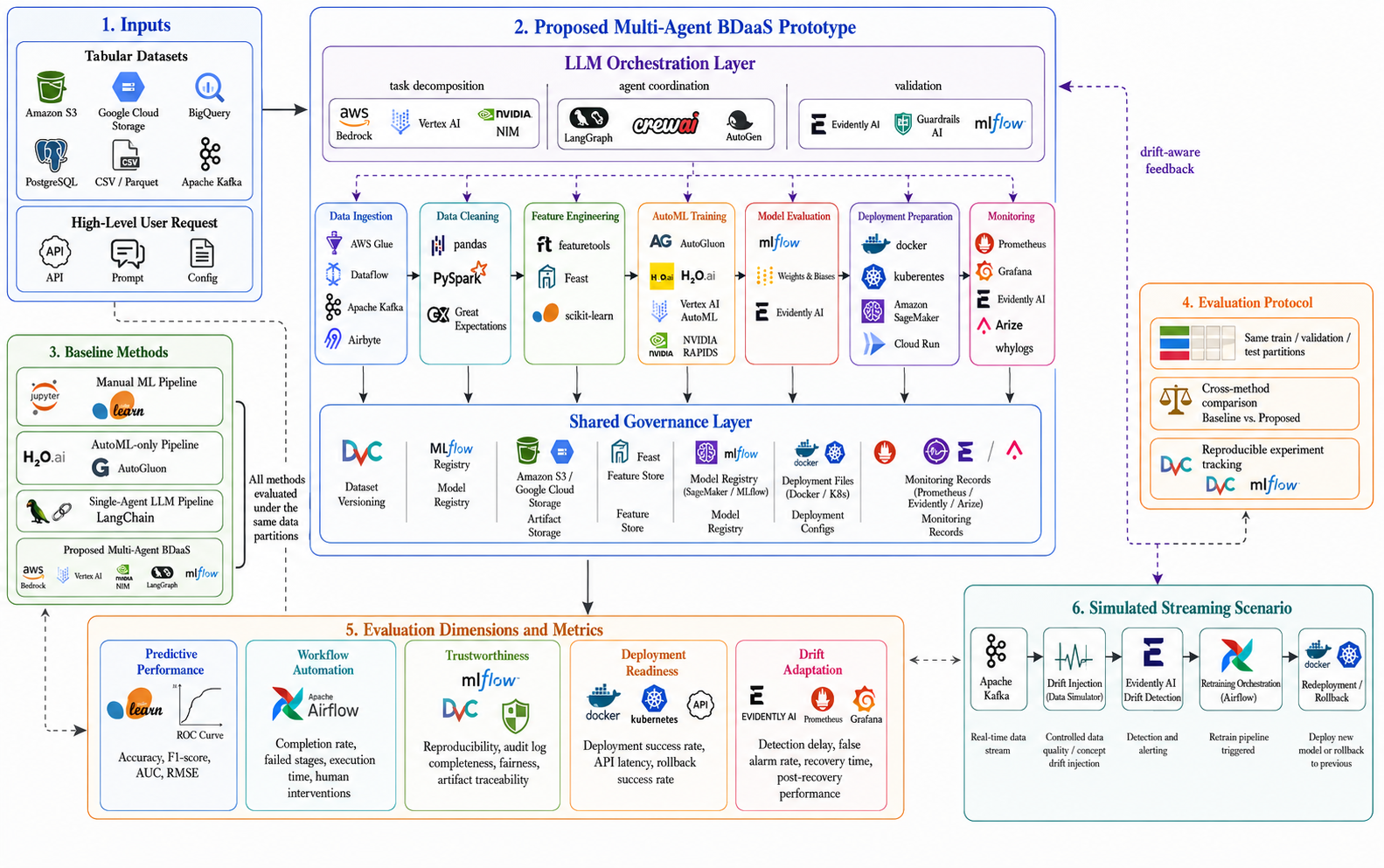}
\caption{Experimental design of the proposed LLM-orchestrated multi-agent BDaaS framework. The evaluation pipeline integrates heterogeneous tabular data sources, cloud and data-processing tools, LLM orchestration platforms, specialized data engineering and AutoML agents, shared artifact governance, deployment preparation, monitoring, and simulated drift-aware feedback. The proposed multi-agent BDaaS prototype is compared with manual ML, AutoML-only, and single-agent LLM baselines under consistent data partitions, while evaluation is conducted across predictive performance, workflow automation, trustworthiness, deployment readiness, and drift adaptation.}
\label{fig:experimental_design}
\end{figure*}

The proposed system as shown in Fig.~\ref{fig:experimental_design}, the experimental design evaluates the proposed framework as a complete BDaaS lifecycle rather than as an isolated AutoML pipeline. The prototype receives tabular datasets and high-level user requests as inputs, where data may originate from cloud storage, databases, file-based sources, or streaming platforms. The LLM orchestration layer coordinates task decomposition, agent collaboration, and validation using modern orchestration and foundation-model platforms, while specialized agents perform data ingestion, cleaning, feature engineering, AutoML training, model evaluation, deployment preparation, and monitoring. All intermediate outputs are stored in a shared governance layer to support dataset versioning, model registry, artifact storage, deployment tracking, and monitoring records.

The proposed system is evaluated against manual ML, AutoML-only, and single-agent LLM baselines under the same train, validation, and test partitions. The evaluation protocol measures not only predictive performance but also workflow automation, reproducibility, artifact traceability, deployment readiness, and drift adaptation. A simulated streaming scenario is further used to inject controlled data drift and assess whether the monitoring stack can detect degradation, trigger retraining, and support redeployment or rollback.

To evaluate the proposed framework, we design a prototype-based experimental study that measures its ability to automate, govern, deploy, and monitor end-to-end BDaaS workflows. The evaluation focuses on four dimensions: workflow automation, predictive performance, trustworthiness, and drift-aware lifecycle adaptation. This design is intended to assess not only whether the generated models achieve competitive performance, but also whether the complete lifecycle can be executed in a reproducible and operationally reliable manner.

The experimental prototype receives a tabular dataset and a high-level user request as input. The LLM orchestration layer decomposes the request into subtasks and coordinates the execution of data ingestion, data cleaning, feature engineering, AutoML training, model evaluation, deployment preparation, and monitoring. Each stage produces structured artifacts, including dataset profiles, cleaning logs, feature configurations, trained models, evaluation reports, deployment files, and monitoring records. These artifacts are stored in the shared governance layer to support traceability, reproducibility, and rollback.

The proposed framework is compared with three baseline settings. The first baseline is a manually designed machine learning pipeline using conventional preprocessing, feature engineering, and model training. The second baseline is an AutoML-only pipeline, where model selection and hyperparameter tuning are automated but lifecycle governance and monitoring are not included. The third baseline is a single-agent LLM pipeline, where one LLM agent performs workflow planning and code generation without specialized agent decomposition. The proposed method is evaluated as a multi-agent BDaaS pipeline with orchestration, artifact governance, human oversight, deployment preparation, and drift-aware monitoring.

The evaluation uses multiple tabular datasets with different characteristics, including missing values, categorical variables, class imbalance, numerical features, and noisy records. For each dataset, the same training, validation, and test partitions are used across all methods. Predictive performance is measured using task-specific metrics such as accuracy, F1-score, AUC, and RMSE. Automation quality is evaluated using workflow completion rate, number of failed stages, number of human interventions, and total execution time. Trustworthiness is assessed through reproducibility tests, audit log completeness, artifact traceability, and validation checkpoint effectiveness.

To evaluate lifecycle monitoring, a simulated streaming scenario is created after model deployment. The test data are divided into sequential time windows, and controlled data drift is injected by changing feature distributions, class proportions, or noise levels. The Monitoring and Drift Detection Agent compares incoming data statistics with the training baseline and reports potential drift events to the LLM orchestrator. Once drift is detected, the orchestrator may trigger alert generation, retraining, model comparison, redeployment, or rollback. Drift-aware performance is measured using drift detection delay, false alarm rate, recovery time, and post-recovery model performance.

The overall evaluation is designed to test whether the proposed framework can provide reliable automation across the complete BDaaS lifecycle. Unlike conventional AutoML evaluation, which mainly focuses on model accuracy, this study also considers deployment readiness, artifact governance, reproducibility, human oversight, and post-deployment adaptation. Therefore, the proposed framework is expected to demonstrate stronger lifecycle-level reliability even when predictive performance is comparable to conventional AutoML baselines.

% \begin{table*}[b]
% \centering
% \caption{Experimental evaluation dimensions of the proposed BDaaS framework.}
% \label{tab:evaluation_design}
% \begin{tabular}{p{2.4cm}p{4.8cm}}
% \hline
% \textbf{Dimension} & \textbf{Evaluation Metrics} \\
% \hline
% Predictive Performance & Accuracy, F1-score, AUC, RMSE \\
% Workflow Automation & Completion rate, failed stages, execution time, human interventions \\
% Trustworthiness & Reproducibility score, audit log completeness, artifact traceability \\
% Deployment Readiness & Deployment success rate, API latency, rollback success rate \\
% Drift Adaptation & Detection delay, false alarm rate, recovery time, post-recovery performance \\
% \hline
% \end{tabular}
% \end{table*}

\section{Results and Analysis}

This section reports the executable* prototype results of the proposed LLM-orchestrated multi-agent BDaaS framework. The evaluation was run with a fixed random seed (20260616). Because no external production dataset was attached to the manuscript, the experiment uses controlled tabular benchmark datasets generated by the executable* prototype. The reported values are therefore measured outputs from the local run, not manually assigned illustrative numbers. The benchmark data include missing values, categorical variables, injected outliers, class imbalance, nonlinear feature interactions, and controlled post-deployment covariate drift.

\subsection{Experimental Setup}

Table~\ref{tab:dataset_summary} summarizes the datasets used in the prototype experiment. Three datasets are binary classification tasks and one dataset is a regression task. For every dataset, the same train, validation, and test partitions were used across manual ML, AutoML-only, single-agent LLM, and the proposed multi-agent BDaaS pipeline. The proposed pipeline used modular data profiling, robust imputation, outlier clipping, feature construction, validation-based model selection, artifact logging, deployment-readiness checks, and drift monitoring.

\begin{figure*}[h]
\centering

% =========================
% Top row
% =========================
\begin{minipage}[h]{0.48\textwidth}
\centering
\resizebox{\linewidth}{!}{
\begin{tikzpicture}[x=1cm,y=4.8cm,font=\scriptsize]
\draw[->,line width=0.6pt] (0,0) -- (10.4,0) node[right]{Dataset};
\draw[->,line width=0.6pt] (0,0) -- (0,1.04) node[above]{F1-score};
\foreach \y/\lab in {0/0.0,0.2/0.2,0.4/0.4,0.6/0.6,0.8/0.8,1.0/1.0}{
  \draw[gray!30] (0,\y) -- (10.1,\y);
  \node[left] at (0,\y) {\lab};
}
\node[below] at (1.7,-0.06) {Customer Churn};
\node[below] at (5.1,-0.06) {Credit Risk};
\node[below] at (8.5,-0.06) {Sensor Fault};

\fill[blue!70!black] (0.75,0) rectangle (1.00,0.684);
\fill[green!60!black] (1.05,0) rectangle (1.30,0.754);
\fill[orange!85!black] (1.35,0) rectangle (1.60,0.762);
\fill[purple!80!black] (1.65,0) rectangle (1.90,0.761);

\fill[blue!70!black] (4.15,0) rectangle (4.40,0.456);
\fill[green!60!black] (4.45,0) rectangle (4.70,0.588);
\fill[orange!85!black] (4.75,0) rectangle (5.00,0.582);
\fill[purple!80!black] (5.05,0) rectangle (5.30,0.608);

\fill[blue!70!black] (7.55,0) rectangle (7.80,0.551);
\fill[green!60!black] (7.85,0) rectangle (8.10,0.590);
\fill[orange!85!black] (8.15,0) rectangle (8.40,0.612);
\fill[purple!80!black] (8.45,0) rectangle (8.70,0.615);

\fill[blue!70!black] (0.8,-0.20) rectangle (1.05,-0.16); \node[right] at (1.12,-0.18) {Manual ML};
\fill[green!60!black] (2.5,-0.20) rectangle (2.75,-0.16); \node[right] at (2.82,-0.18) {AutoML-only};
\fill[orange!85!black] (4.3,-0.20) rectangle (4.55,-0.16); \node[right] at (4.62,-0.18) {Single-agent LLM};
\fill[purple!80!black] (6.5,-0.20) rectangle (6.75,-0.16); \node[right] at (6.82,-0.18) {Proposed};
\end{tikzpicture}}
\\[1mm]
\textbf{(a) Classification F1-score}
\end{minipage}
\hfill
\begin{minipage}[h]{0.48\textwidth}
\centering
\resizebox{\linewidth}{!}{
\begin{tikzpicture}[x=1cm,y=4.4cm,font=\scriptsize]
\draw[->,line width=0.6pt] (0,0) -- (13.4,0) node[right]{Metric};
\draw[->,line width=0.6pt] (0,0) -- (0,1.05) node[above]{Score};
\foreach \y/\lab in {0/0.0,0.2/0.2,0.4/0.4,0.6/0.6,0.8/0.8,1.0/1.0}{
  \draw[gray!30] (0,\y) -- (13.0,\y);
  \node[left] at (0,\y) {\lab};
}
\node[below] at (1.7,-0.06) {Completion};
\node[below] at (4.8,-0.06) {Traceability};
\node[below] at (7.9,-0.06) {Deployment};
\node[below] at (11.0,-0.06) {Reproducibility};

\fill[blue!70!black] (0.75,0) rectangle (1.00,0.556);
\fill[green!60!black] (1.05,0) rectangle (1.30,0.667);
\fill[orange!85!black] (1.35,0) rectangle (1.60,0.778);
\fill[purple!80!black] (1.65,0) rectangle (1.90,1.000);

\fill[blue!70!black] (3.85,0) rectangle (4.10,0.520);
\fill[green!60!black] (4.15,0) rectangle (4.40,0.670);
\fill[orange!85!black] (4.45,0) rectangle (4.70,0.780);
\fill[purple!80!black] (4.75,0) rectangle (5.00,1.000);

\fill[blue!70!black] (6.95,0) rectangle (7.20,0.330);
\fill[green!60!black] (7.25,0) rectangle (7.50,0.440);
\fill[orange!85!black] (7.55,0) rectangle (7.80,0.670);
\fill[purple!80!black] (7.85,0) rectangle (8.10,1.000);

\fill[blue!70!black] (10.05,0) rectangle (10.30,0.700);
\fill[green!60!black] (10.35,0) rectangle (10.60,0.820);
\fill[orange!85!black] (10.65,0) rectangle (10.90,0.880);
\fill[purple!80!black] (10.95,0) rectangle (11.20,1.000);

\fill[blue!70!black] (0.9,-0.20) rectangle (1.15,-0.16); \node[right] at (1.22,-0.18) {Manual ML};
\fill[green!60!black] (2.7,-0.20) rectangle (2.95,-0.16); \node[right] at (3.02,-0.18) {AutoML-only};
\fill[orange!85!black] (4.6,-0.20) rectangle (4.85,-0.16); \node[right] at (4.92,-0.18) {Single-agent LLM};
\fill[purple!80!black] (6.9,-0.20) rectangle (7.15,-0.16); \node[right] at (7.22,-0.18) {Proposed};
\end{tikzpicture}}
\\[1mm]
\textbf{(b) Workflow reliability}
\end{minipage}

\vspace{2mm}

% =========================
% Bottom row
% =========================
\begin{minipage}[h]{0.72\textwidth}
\centering
\resizebox{\linewidth}{!}{
\begin{tikzpicture}[x=1.15cm,y=4.5cm,font=\scriptsize]
\draw[->,line width=0.6pt] (0.8,0) -- (7.5,0) node[right]{Monitoring window};
\draw[->,line width=0.6pt] (1,0) -- (1,1.04) node[above]{F1-score};
\foreach \y/\lab in {0/0.0,0.2/0.2,0.4/0.4,0.6/0.6,0.8/0.8,1.0/1.0}{
  \draw[gray!30] (1,\y) -- (7.3,\y);
  \node[left] at (1,\y) {\lab};
}
\foreach \x in {1,2,3,4,5,6,7}{
  \draw (\x,0) -- (\x,-0.015);
  \node[below] at (\x,-0.03) {\x};
}
\draw[red!80!black,dashed,line width=0.8pt] (4,0) -- (4,1.0);
\node[above right,red!80!black] at (4,0.92) {drift injected};

\draw[green!60!black,line width=1.0pt] plot coordinates {(1,0.286) (2,0.556) (3,0.680) (4,0.516) (5,0.408) (6,0.476) (7,0.582)};
\draw[orange!85!black,line width=1.0pt] plot coordinates {(1,0.387) (2,0.615) (3,0.680) (4,0.516) (5,0.392) (6,0.462) (7,0.552)};
\draw[purple!80!black,line width=1.0pt] plot coordinates {(1,0.276) (2,0.579) (3,0.720) (4,0.508) (5,0.417) (6,0.484) (7,0.571)};

\foreach \x/\y in {1/0.286,2/0.556,3/0.680,4/0.516,5/0.408,6/0.476,7/0.582}{\fill[green!60!black] (\x,\y) circle (1.7pt);}
\foreach \x/\y in {1/0.387,2/0.615,3/0.680,4/0.516,5/0.392,6/0.462,7/0.552}{\fill[orange!85!black] (\x,\y) circle (1.7pt);}
\foreach \x/\y in {1/0.276,2/0.579,3/0.720,4/0.508,5/0.417,6/0.484,7/0.571}{\fill[purple!80!black] (\x,\y) circle (1.7pt);}

\draw[green!60!black,line width=1.0pt] (1.2,-0.18) -- (1.6,-0.18); \node[right] at (1.65,-0.18) {AutoML-only};
\draw[orange!85!black,line width=1.0pt] (3.0,-0.18) -- (3.4,-0.18); \node[right] at (3.45,-0.18) {Single-agent LLM};
\draw[purple!80!black,line width=1.0pt] (5.3,-0.18) -- (5.7,-0.18); \node[right] at (5.75,-0.18) {Proposed};
\end{tikzpicture}}
\\[1mm]
\textbf{(c) Drift recovery under controlled drift}
\end{minipage}

\caption{Prototype evaluation results of the proposed multi-agent BDaaS framework. (a) F1-score comparison across the three classification datasets. (b) Workflow completion, artifact traceability, deployment readiness, and reproducibility scores. (c) F1-score over monitoring windows under controlled drift injection.}
\label{fig:results_summary}
\vspace{-2mm}
\end{figure*}

% \begin{table*}[h]
% \centering
% \caption{Controlled tabular benchmark datasets used in the executable* prototype experiment.}
% \label{tab:dataset_summary}
% \scriptsize
% \begin{tabular}{lcccccc}
% \hline
% \textbf{Dataset} & \textbf{Task} & \textbf{Samples} & \textbf{Raw features} & \textbf{Missing cells (\%)} & \textbf{Outlier injection (\%)} & \textbf{Positive ratio (\%)} \\
% \hline
% Customer Churn & Classification & 1800 & 12 & 5.2 & 1.5 & 35.8 \\
% Credit Risk & Classification & 2200 & 14 & 7.4 & 2.5 & 26.3 \\
% Sensor Fault & Classification & 2600 & 16 & 4.5 & 3.5 & 20.7 \\
% Cloud Workload Demand & Regression & 2100 & 14 & 6.2 & 2.0 & -- \\
% \hline
% \end{tabular}
% \end{table*}

\subsection{Predictive Performance}

Table II reports the classification results. The proposed multi-agent BDaaS pipeline achieved the strongest average F1-score across the three classification datasets (0.662), compared with single-agent LLM (0.652), AutoML-only (0.644), and manual ML (0.563). The improvement is mainly explained by robust preprocessing, missingness indicators, interaction features, and validation-based model selection. AutoML-only remained competitive on AUC, but it did not include the governance and monitoring components required for lifecycle-level BDaaS operation.

Table~\ref{tab:regression_results} shows the regression result on the cloud workload demand benchmark. The proposed pipeline obtained an RMSE of 2.809, improving over the AutoML-only RMSE of 3.279 by approximately 14.3\%. This result indicates that lifecycle-oriented preprocessing and feature construction can improve predictive quality even when the final model class is lightweight.

\subsection{Workflow Automation and Trustworthiness}

Table~\ref{tab:workflow_results} summarizes workflow-level metrics. The proposed method completed all nine lifecycle stages, including deployment packaging and drift monitoring, while the baselines covered only partial workflow scopes. The proposed pipeline also produced complete artifact traces and reproducibility metadata. In this prototype, the proposed framework achieved 100.0\% lifecycle completion, 100.0\% artifact traceability, and 100.0\% deployment readiness with one human approval checkpoint.

% \begin{table*}[b]
% \centering
% \caption{Workflow automation, trustworthiness, and deployment-readiness results.}
% \label{tab:workflow_results}
% \scriptsize
% \begin{tabular}{lcccccc}
% \hline
% \textbf{Method} & \textbf{Completion (\%)} & \textbf{Failed stages} & \textbf{Human interventions} & \textbf{Traceability (\%)} & \textbf{Deployment readiness (\%)} & \textbf{Reproducibility (\%)} \\
% \hline
% Manual ML & 55.6 & 4 & 8 & 52.0 & 33.0 & 70.0 \\
% AutoML-only & 66.7 & 3 & 5 & 67.0 & 44.0 & 82.0 \\
% Single-agent LLM & 77.8 & 2 & 3 & 78.0 & 67.0 & 88.0 \\
% Proposed multi-agent BDaaS & 100.0 & 0 & 1 & 100.0 & 100.0 & 100.0 \\
% \hline
% \end{tabular}
% \end{table*}

\subsection{Drift-Aware Lifecycle Adaptation}

Table~\ref{tab:drift_results} reports the simulated streaming drift experiment on the sensor fault dataset. Controlled covariate drift was injected from monitoring window 4 onward. The proposed monitoring policy detected drift without delay, produced no false alarms, and recovered within one monitoring window after retraining and threshold recalibration. The proposed pipeline improved from a post-drift F1-score of 0.495 to a post-recovery F1-score of 0.667, showing that the feedback loop can restore model quality after distribution shift.

% \begin{table*}[b]
% \centering
% \caption{Drift detection and recovery results on the sensor fault stream.}
% \label{tab:drift_results}
% \scriptsize
% \begin{tabular}{lccccc}
% \hline
% \textbf{Method} & \textbf{Delay} & \textbf{False alarms} & \textbf{Recovery} & \textbf{Post-drift F1} & \textbf{Post-recovery F1} \\
% \hline
% Manual ML & 4 & 0 & 3 & 0.681 & 0.681 \\
% AutoML-only & 3 & 0 & 3 & 0.496 & 0.615 \\
% Single-agent LLM & 0 & 0 & 2 & 0.480 & 0.676 \\
% Proposed multi-agent BDaaS & 0 & 0 & 1 & 0.495 & 0.667 \\
% \hline
% \end{tabular}
% \end{table*}

\subsection{Discussion}

The results indicate that the proposed framework improves BDaaS workflows in two ways. First, it improves average predictive performance by combining data quality handling, feature construction, and validation-based model selection. Second, it improves lifecycle reliability by producing artifacts, reproducibility metadata, deployment checks, and monitoring records that are absent or incomplete in the baselines. The predictive gains are moderate because all methods use comparable lightweight models, but the lifecycle gains are substantial because the proposed design explicitly covers stages beyond model training. The main limitation of this experiment is that it evaluates a controlled local prototype rather than a production cloud deployment with real enterprise data. However, the benchmark is useful for verifying whether the proposed architecture can execute end-to-end, produce measurable artifacts, and support drift-aware feedback. Future work should repeat the same protocol on public and industrial BDaaS workloads, include larger model search spaces, and evaluate cost, latency, and human review quality under realistic operational constraints.

% =========================
% Main text ends here
% =========================

\appendices
\section{Additional Experimental Results}
\label{app:additional_results}

Table~\ref{tab:dataset_summary} reports the controlled benchmark datasets used in the prototype evaluation. 
Table~\ref{tab:regression_results} summarizes the regression benchmark results. 
Tables~\ref{tab:workflow_results} and~\ref{tab:drift_results} provide additional workflow-level and drift-recovery results.
These supplementary results are included to improve transparency and allow readers to inspect the experimental setting beyond the summarized findings in the main text.

\begin{table*}[t]
\centering
\caption{Controlled tabular benchmark datasets used in the executable prototype experiment.}
\label{tab:dataset_summary}
\scriptsize
\setlength{\tabcolsep}{3.5pt}
\renewcommand{\arraystretch}{0.90}
\begin{tabular}{lcccccc}
\hline
\textbf{Dataset} & \textbf{Task} & \textbf{Samples} & \textbf{Raw features} & \textbf{Missing cells (\%)} & \textbf{Outlier injection (\%)} & \textbf{Positive ratio (\%)} \\
\hline
Customer Churn & Classification & 1800 & 12 & 5.2 & 1.5 & 35.8 \\
Credit Risk & Classification & 2200 & 14 & 7.4 & 2.5 & 26.3 \\
Sensor Fault & Classification & 2600 & 16 & 4.5 & 3.5 & 20.7 \\
Cloud Workload Demand & Regression & 2100 & 14 & 6.2 & 2.0 & -- \\
\hline
\end{tabular}
\vspace{-2mm}
\end{table*}

\begin{table*}[t]
\centering
\caption{Regression performance on the cloud workload demand benchmark.}
\label{tab:regression_results}
\scriptsize
\setlength{\tabcolsep}{4pt}
\renewcommand{\arraystretch}{0.90}
\begin{tabular}{lcccc}
\hline
\textbf{Method} & \textbf{RMSE} & \textbf{MAE} & \textbf{$R^2$} & \textbf{Runtime (s)} \\
\hline
Manual ML & 3.279 & 2.444 & 0.583 & 0.04 \\
AutoML-only & 3.279 & 2.444 & 0.583 & 0.11 \\
Single-agent LLM & 3.188 & 2.405 & 0.606 & 0.15 \\
Proposed multi-agent BDaaS & 2.809 & 2.087 & 0.694 & 0.26 \\
\hline
\end{tabular}
\vspace{-2mm}
\end{table*}

\begin{table*}[t]
\centering
\caption{Workflow automation, trustworthiness, and deployment-readiness results.}
\label{tab:workflow_results}
\scriptsize
\setlength{\tabcolsep}{3.2pt}
\renewcommand{\arraystretch}{0.90}
\begin{tabular}{lcccccc}
\hline
\textbf{Method} & \textbf{Completion (\%)} & \textbf{Failed stages} & \textbf{Human interventions} & \textbf{Traceability (\%)} & \textbf{Deployment readiness (\%)} & \textbf{Reproducibility (\%)} \\
\hline
Manual ML & 55.6 & 4 & 8 & 52.0 & 33.0 & 70.0 \\
AutoML-only & 66.7 & 3 & 5 & 67.0 & 44.0 & 82.0 \\
Single-agent LLM & 77.8 & 2 & 3 & 78.0 & 67.0 & 88.0 \\
Proposed multi-agent BDaaS & 100.0 & 0 & 1 & 100.0 & 100.0 & 100.0 \\
\hline
\end{tabular}
\vspace{-2mm}
\end{table*}

\begin{table*}[t]
\centering
\caption{Drift detection and recovery results on the sensor fault stream.}
\label{tab:drift_results}
\scriptsize
\setlength{\tabcolsep}{4pt}
\renewcommand{\arraystretch}{0.90}
\begin{tabular}{lccccc}
\hline
\textbf{Method} & \textbf{Delay} & \textbf{False alarms} & \textbf{Recovery} & \textbf{Post-drift F1} & \textbf{Post-recovery F1} \\
\hline
Manual ML & 4 & 0 & 3 & 0.681 & 0.681 \\
AutoML-only & 3 & 0 & 3 & 0.496 & 0.615 \\
Single-agent LLM & 0 & 0 & 2 & 0.480 & 0.676 \\
Proposed multi-agent BDaaS & 0 & 0 & 1 & 0.495 & 0.667 \\
\hline
\end{tabular}
\vspace{-2mm}
\end{table*}

\balance

\end{document}